\newif\ifarxiv
\arxivtrue
\documentclass{article}

\usepackage{cite}
\usepackage{amsmath,amssymb,amsfonts}
\usepackage{graphicx}
\usepackage{textcomp}
\usepackage{xcolor}

\ifarxiv
\usepackage[paper=a4paper,margin=2.3cm,bottom=3cm]{geometry}
\usepackage{hyperref}
\makeatletter
\renewenvironment{table}
               {\def\@floatboxreset{\reset@font\scriptsize\@setminipage}\@float{table}}
               {\end@float}
\renewenvironment{table*}
               {\def\@floatboxreset{\reset@font\scriptsize\@setminipage}\@dblfloat{table}}
               {\end@dblfloat}
\makeatother
\else
\fi
\usepackage{ifthen}
\input{dgfly_drawer}

\usepackage{pgfplots}
\usepgfplotslibrary{statistics}
\pgfplotsset{compat=newest}
\pgfplotsset{minor grid style={dashed,very thin, color=blue!15}}
\pgfplotsset{major grid style={very thin, color=black!30}}

\pgfkeys{/pgf/number format/.cd,
fixed,
fixed zerofill=false,
set thousands separator={\,},
1000 sep in fractionals,
sci E,
}

\newenvironment{experimentfigure}{\begin{figure*}}{\end{figure*}}

\makeatletter
\tikzset{nomorepostaction/.code=\let\tikz@postactions\pgfutil@empty}
\makeatother

\pgfplotsset{
	automatically generated axis/.style={
		height=105pt,
		width=174pt,
		scaled ticks=false,
		xticklabel style={font=\tiny,/pgf/number format/.cd, fixed,/tikz/.cd},
		yticklabel style={font=\tiny,/pgf/number format/.cd, fixed,/tikz/.cd},
		x label style={at={(ticklabel cs:0.5, -5pt)},name={x label},anchor=north,font=\scriptsize},
		y label style={at={(ticklabel cs:0.5, -5pt)},name={y label},anchor=south,font=\scriptsize},
		title style={font=\scriptsize,at={(0.5,0.95)},anchor=base},
		ymin=0,ymax=0.51,xmin=0,%
		ytick={0,0.1,...,0.5},
	},
	automatically generated symbolic/.style={
		height=105pt,
		width=500pt,
		xticklabel style={font=\tiny,rotate=90},
		yticklabel style={font=\tiny,/pgf/number format/.cd, fixed,/tikz/.cd},
		x label style={at={(ticklabel cs:0.5, -5pt)},name={x label},anchor=north,font=\scriptsize},
		y label style={at={(ticklabel cs:0.5, -5pt)},name={y label},anchor=south,font=\scriptsize},
	},
	first kind/.style={
		legend style={font=\scriptsize,fill=none},
		legend columns=4,legend cell align=left,
	},
	posterior kind/.style={
		legend style={draw=none},
	},
}
\tikzset{
	automatically generated plot/.style={
		/pgfplots/error bars/error bar style={ultra thin,solid,opacity=0.25},
		/tikz/mark options={solid},
	},
	automatically generated bar plot/.style={
	},
	automatically generated boxplot/.style={
	},
	temporal plot/.style={
		mark repeat=100,
		mark phase=5,
		/pgfplots/error bars/x dir=none,
		/pgfplots/error bars/y dir=none,
	},
}

\tikzset{xValiantxoverxminAx/.style={automatically generated plot,red,solid,mark=diamond,every mark/.append style={color=red!30!black}}}
\tikzset{xValiantxoverxminBx/.style={automatically generated plot,olive,solid,mark=o,every mark/.append style={color=olive!30!black}}}
\tikzset{xValiantxoverxDORxminAx/.style={xValiantxoverxminAx}}
\tikzset{xValiantxoverxDORxminBx/.style={xValiantxoverxminBx}}
\tikzset{xValiantxoverxladderx/.style={automatically generated plot,blue,solid,mark=triangle,every mark/.append style={color=blue!30!black}}}
\tikzset{xValiantxoverxladderxreusex/.style={automatically generated plot,black,solid,mark=Mercedes star flipped}}
\tikzset{seed/.style={opacity=0.5,very thin}}

\def\xValiantxoverxDORxminAxtext{Valiant 2phases MinFirst}
\def\xValiantxoverxDORxminBxtext{Valiant 2phases MinLast}
\def\xValiantxoverxladderxtext{Valiant Ladder}
\def\xValiantxoverxladderxreusextext{Valiant Ladder reuse}
\def\xValiantxoverxminAxtext{Valiant 2phases MinFirst}
\def\xValiantxoverxminBxtext{Valiant 2phases MinLast}

\usepackage[utf8]{inputenc}
\usetikzlibrary{positioning}
\usetikzlibrary{decorations.text}
\usetikzlibrary{calc}
\usetikzlibrary{arrows, decorations.markings}
\usetikzlibrary{matrix}
\usepackage[normalem]{ulem}
\usepackage{amsthm}

\usetikzlibrary{patterns}
\usetikzlibrary{external}
\tikzexternaldisable
\ifnum\pdfshellescape=1
\else
\fi

\def\SWITCH#1{\textsl{#1}}
\def\SS{\SWITCH{SS}}
\def\IS{\SWITCH{IS}}
\def\DS{\SWITCH{DS}}

\begin{document}


\ifarxiv
\title{Analysing Mechanisms for Virtual Channel Management in Low-Diameter networks}
\fi

\ifarxiv
\author{Alejandro Cano \and Cristóbal Camarero \and Carmen Martínez \and Ramón Beivide}
\fi

\maketitle

\begin{abstract}
To interconnect their growing number of servers, current supercomputers and data centers are starting to adopt low-diameter networks, such as HyperX, Dragonfly and Dragonfly+. These emergent topologies require balancing the load over their links and finding suitable non-minimal routing mechanisms for them becomes particularly challenging. The Valiant load balancing scheme is a very popular choice for non-minimal routing. Evolved adaptive routing mechanisms implemented in real systems are based on this Valiant scheme.

All these low-diameter networks are deadlock-prone when non-minimal routing is employed. Routing deadlocks occur when packets cannot progress due to cyclic dependencies. Therefore, developing efficient deadlock-free packet routing mechanisms is critical for the progress of these emergent networks. The routing function includes the routing algorithm for path selection and the buffers management policy that dictates how packets allocate the buffers of the switches on their paths. For the same routing algorithm, a different buffer management mechanism can lead to a very different performance. Moreover, certain mechanisms considered efficient for avoiding deadlocks, may still suffer from hard to pinpoint instabilities that make erratic the network response. This paper focuses on exploring the impact of these buffers management policies on the performance of current interconnection networks, showing a 90\% of performance drop if an incorrect buffers management policy is used. Moreover, this study not only characterizes some of these undesirable scenarios but also proposes practicable solutions.
\end{abstract}

\section{Introduction}\label{sec:intro}

\begin{figure*}
    \centering%
    \tikzsetnextfilename{externalized-figure-topologies}%
    \begin{tikzpicture}
        \node at (0,0) {\dgflydrawrect{4}{4}{3}{4}{hamming}{.2\paperwidth}};
        \node at (6,0) {\dgflydraw{4}{9}{2}{1}{palmtree}{.25\paperwidth}};
        \node at (12,0) {\dgflyplusdraw{3}{10}{3}{1}{palmtree}{.25\paperwidth}};
    \end{tikzpicture}
    \caption{Small instances of the studied topologies: 2D HyperX, Dragonfly and Dragonfly+ respectively. Switches are represented as solid rectangles and servers are omitted. The links coming out of a selected group are in bold.}
    \label{fig:topos_used}
\end{figure*}
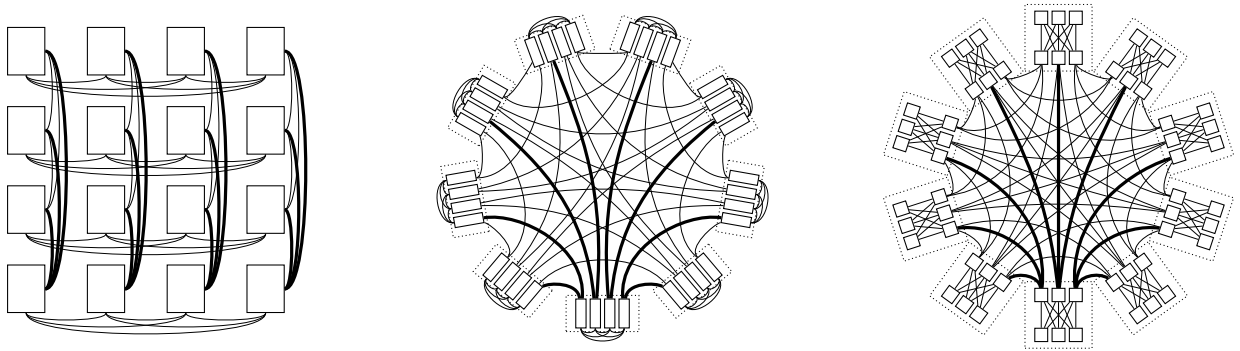

The \textit{diameter} is an important attribute of a network, which is the maximum distance between any pair of servers when using minimal routing. Current supercomputers and datacenters are starting to use low-diameter networks for interconnecting their increasing number of servers.
Modern networks such as Dragonflies~\cite{Kim_dgfly_ISCA}, 3D HyperX~\cite{HyperX} (or 3D Flattened Butterflies~\cite{kim_flat_CAL}) and Dragonfly+~\cite{dragonfly_plus} (or Megafly~\cite{megafly}) have diameter 3, and have been used in different machines during the last decade. As these low-diameter networks are cheaper than Fat-trees, assuming adequate routing mechanisms are provided, a higher market penetration is expected for them.

Networks are modelled by graphs, with vertices representing switches and edges representing links. Thus, the diameter is the number of hops of the longest minimal path. For example, in a \textit{complete} graph of $v$ vertices, $K_{v}$, every pair of vertices is connected by an edge, which leads to a network with diameter one, $D=1$, and $v(v-1)/2\approx v^{2}/2$ edges. Dragonfly, HyperX, and Dragonfly+ networks are based on complete graphs. Figure~\ref{fig:topos_used} shows small instances of these topologies.

The Dragonfly is a \textit{global} complete graph connecting super-nodes or groups of switches, which, in turn, are connected by means of \textit{local} complete graphs. Thus, it is a two-level hierarchical network with $D=3$. The Frontier supercomputer, currently number one in the TOP500 list~\cite{Top500}, employs a Dragonfly of this type. The Dragonfly+ network is similar to the Dragonfly but, for implementing its groups, it employs leaf/spine Fat-trees instead of complete graphs. As it uses a \textit{global} complete graph, it is also a diameter 3 topology, but scales better than Dragonflies. There are currently few machines using the Dragonfly+, but it is expected a higher market share for them. In the last TOP500 list, the Leonardo supercomputer~\cite{Leonardo} is in 4th position and employs a Dragonfly+. HyperX networks are Cartesian graph powers of complete graphs. A HyperX of $n$ dimensions and $v$ vertices per dimension is modeled by the $K_{v}^{n}$ graph. The diameter of these HyperX networks is precisely $n$, their number of dimensions. HyperX networks of two and three dimensions have been employed in some deployments~\cite{FT-hyperx,Cascade}.

A good topology is useless without a proper routing. Finding efficient schemes to convey packets is specially challenging in these low-diameter networks, as non-minimal routing is compulsory for them. In these networks, it is known that minimal paths must be selected when the traffic is benign (uniform) and non-minimal, up to doubling the length of minimal paths, when the traffic is adverse (biased). The Valiant~\cite{Valiant_SIAM} load balancing scheme is one of the most popular non-minimal routing mechanisms to deal with adverse traffic. Different evolved routing mechanisms, such as UGAL~\cite{Singh}, are based on it. With Valiant, before injecting a packet in the network, a random intermediate switch, \IS{}, is selected. Then, the packet traverses the network in two phases, or what is the same, using paths of two segments: first, it is minimally sent from the source switch, \SS{}, to \IS{}, and then, from \IS{} to the destination switch, \DS{}, following again a minimal path.

The previous networks are all deadlock-prone when using non-minimal routing. Routing deadlocks, in which a set of packets cannot advance because of cyclic dependencies among them, are catastrophic situations that can provoke a system down. Thus, the design of efficient deadlock-free routing algorithms is of paramount importance. Routing deadlock is typically avoided either by restricting the paths to use, or by restricting packet injection, or by adding \textit{buffer classes} (or \textit{virtual channels}, VCs) to the switch ports and restricting their use. In commercial systems, it is common to find different mixes of these strategies.  Moreover, virtual channels are typically added to switch ports not just for breaking deadlocks but for improving performance by alleviating some head-of-line (HoL) blocking. HoLB is a performance-limiting phenomenon that occurs when a line of packets is held up in a FIFO queue by the first packet.

When using minimal paths, Dimensionally Ordered Routing (DOR) is a popular deadlock-free mechanism that restricts network routes and can be applied to Meshes and other Cartesian topologies. The deadlock-free Up/Down routing, used in Fat-trees, also belongs to this path-restricted class. Bubble flow control~\cite{Carrion} for Rings and Tori, belongs instead, to the class of injection restriction, but Dateline~\cite{dateline} restricts the use of buffer classes or VCs.

Each deadlock-free routing solution comes with a different cost/performance ratio. Buffers are precious resources of switches, which condition cost, as they are power hungry and require a lot of silicon area. Many current switches employ buffers located at both input and output ports. The area devoted to them scales with the number of ports of the switch, or \textit{radix}, and current switches employ 64 ports, or even more. On top of that, for avoiding protocol deadlocks and/or providing quality of service, the number of buffers needed per port must be multiplied by the number of packet classes. Hence, it has been typical to look for deadlock-free routing mechanisms that do not require too many buffer classes or VCs. For example, DOR does not require buffer classes to avoid deadlocks with minimal routing, but exhibits poor performance under non-uniform traffics.

This paper presents an in-depth analysis of the impact that different buffer management strategies have on the performance of current low-diameter interconnection networks. The main contributions of this work are:
\begin{itemize}
	\item Identify the conditions for which the Valiant mechanism, traditionally considered independent of the traffic pattern, may become unstable to the point of being unusable.
	\item To evaluate current topologies in terms of their susceptibility to unstable behaviour.
	\item To identify VC management policies in which small design changes lead to whole different behaviors under some traffic patterns.
	\item To explain some anomalies in previous results found in the literature.
	\item To show possible trade-offs between raw potential throughput and network response stability.
    \item To provide insights and recommendations for the design of efficient routing mechanisms.
\end{itemize}

The remainder of the paper is organized as follows.
Section~\ref{sec:deadlock} describes the main deadlock avoidance solutions when using Valiant routing in low-diameter networks.
Section~\ref{sec:decoupling} considers these deadlock avoidance solutions as decoupled from routing, and shows how they have been used in other works.
Section~\ref{sec:setup} outlines the experimental setup and methodology employed. It describes the simulation environment, the chosen network topologies, the traffic patterns utilized, and the specific evaluation used to assess the performance of the VC management mechanisms.
Section~\ref{sec:results}  presents detailed results and analysis obtained from the experimental evaluations. It provides insights into the performance, stability, and throughput of the distinct VC policies in the different networks and under different traffic patterns.
Section~\ref{sec:conclusions} summarizes the key findings of the study and draws conclusions based on the obtained results.

%

\section{Deadlock Avoidance Mechanisms}\label{sec:deadlock}
    This section classifies and describes the principal deadlock avoidance mechanisms that can be employed to support non-minimal Valiant routing in low-diameter topologies.

    \subsection{VC Management in Two Phases}

    Similarly to other Cartesian topologies, the low-cost DOR mechanism can be applied to HyperX networks (henceforth written HX) for avoiding deadlocks when using minimal routing. This is, in principle, one of the reasons for which these topologies are appealing. Nevertheless, DOR is not enough to break packet deadlocks when using Valiant routing. Remember that Valiant has paths composed of two segments delimited by the intermediate switch \IS{}. Then, in a natural way, two VCs, one devoted to the first phase of the route and the other to the second, are enough to guaranty deadlock freedom with Valiant in a HX. We denote such a VC management policy as \textit{2Phases}. Observe that the minimum number of required VCs, which is 2, does not depend on the number of network dimensions and, thus, is independent of the length of the traversed paths.

    Similarly, minimal routing in a Dragonfly+ network (DF+) is deadlock-free, and in the same way, two VCs, one for each phase, are needed to implement a deadlock-free Valiant load balancing scheme.
    
With respect to the Dragonfly (DF), the minimal routes are those that minimize the number of global links, this is, minimal routes including two or more global links are forbidden. Thus, the minimal paths in DFs are of type \textit{local-global-local} (henceforth written \texttt{lgl}), which require 2 VCs on local ports, visited in order, to avoid deadlock with minimal routing.
These paths start with a local link inside the source group, then a global link to achieve the destination group, and finally a local link to arrive to the destination switch; with any of the links being possibly omitted depending on the relative locations of the source and destination switches.
	Thus, for implementing Valiant, 2 local ordered VCs and 1 global VC is required per phase, employing a total of 4 and 2 VCs, for local and global ports, respectively.

    Although one VC is enough per Valiant segment in HX and DF+ networks, or two VCs for the DF case, in 2Phases it is common to use more VCs per phase to increase performance by reducing HoL blocking, which keeps links busier. When using more than one VC per phase, a JSQ (join shortest queue) discipline is commonly applied.

    \subsection{Ladder VC Management} 

	An old method that employs VCs to break deadlock is to increase the order of the visited VC with each link traversed by the packet~\cite{Gunter, Merlin}.
	When minimal routing is used on a network of diameter $D$, this method requires $D$ virtual channels per port. Packets traversing longest routes are injected at $VC_{0}$ and ejected at $VC_{D-1}$, visiting all the VCs on their path and incrementing the VC order after every hop. Other packets traveling by shorter routes are also injected at $VC_{0}$ and travel in the same staggered way, but are ejected earlier. This buffer management strategy is denoted as \textit{Ladder} in this paper, and a hop is a \textit{step} of this ladder. Many classic topologies, as Meshes and Tori, have large diameter and cannot use a Ladder mechanism due to its high cost. Nevertheless, the recent advent of low-diameter networks recovered the interest on it. For example, it has been recently used in~\cite{Kim_omni} for HX.

    With the Ladder scheme, independently of the current Valiant phase, the $i$th-hop of a packet goes through the $VC_{i}$ of a certain port of a certain switch in the network. The longest paths in Valiant comprises up to $2D$ hops, so a Ladder for this routing will have $2D$ steps or VCs. As $D=n$ in a HX, the cost of a Ladder, in terms of VCs, would be $2n$. Both DF and DF+ have $D=3$, so with Valiant a packet could take up to six hops, requiring six steps (or VCs) in the Ladder.

	To increase the Ladder performance, more VCs can be allocated to each step. This is done in the same way as with 2Phases, using a JSQ discipline.

    Variants of the Ladder scheme, dealing separately with local and global switch ports, are in use in industrial products for DFs. Actually, the previous 2Phases VC manager in a DF can be seen as two separate Ladders, one having 4 steps for local hops and the other having 2, for the global ones.

\begin{figure}
    \centering%
    \hskip-1ex\includegraphics[scale=0.45]{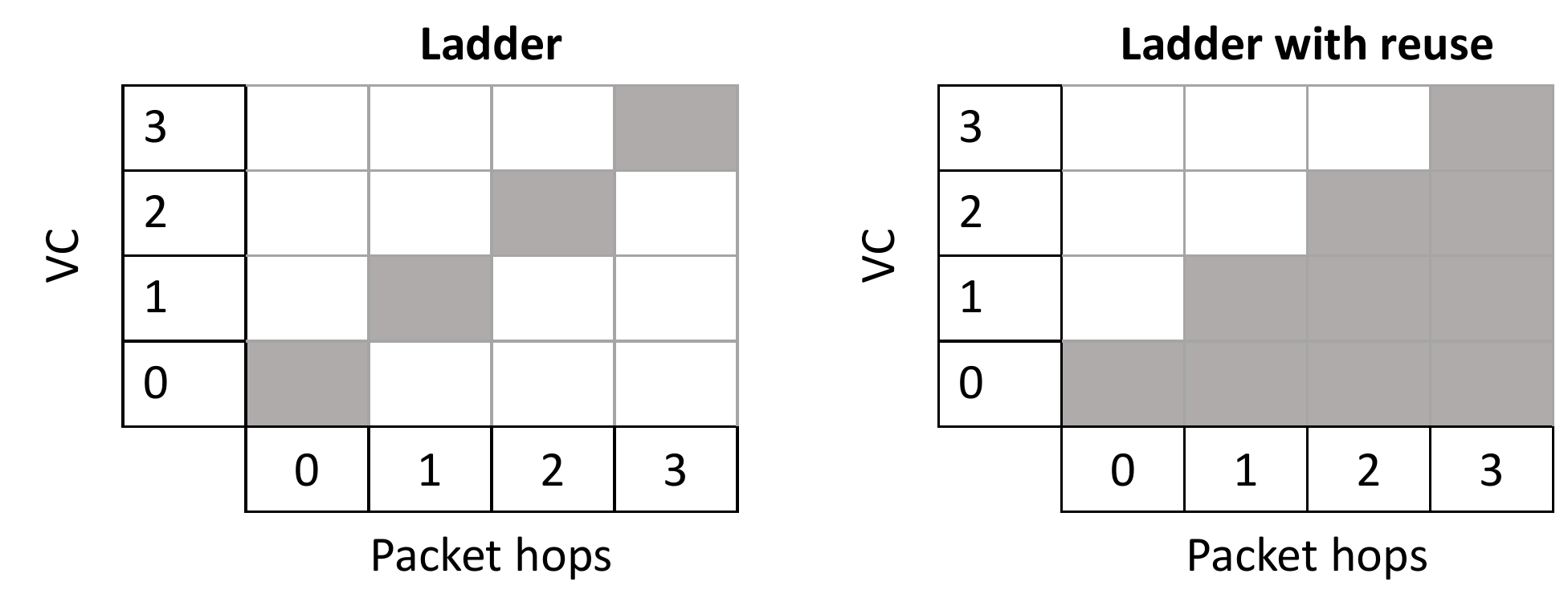}
    \caption{Available VCs (in gray) for a packet that has traveled a given number of hops in Ladder (left) and Ladder with reused VCs (right).}
	\label{fig:ladder-vcs}
\end{figure}

The Ladder mechanism can be optimized to increment the utilization of VCs and reduce HoLB, by allowing to use more than one VC at several hops of the packet, as done for example in \cite{Giessler,Marina-OLM}. Therefore, we define a variant of the Ladder scheme to act as the following: at every $i$th-hop, not only the $i$th-buffer is allowed, but also any $j$th-buffer constrained by ${0 \leq j \leq i}$. We denote to this scheme as \textit{Ladder with reused VCs}. Both Ladder and Ladder with reused VCs are illustrated in Figure~\ref{fig:ladder-vcs}.

\section{Decoupling VC Management from Routing}\label{sec:decoupling}


    In networks that use VCs, it is useful to consider the route generation and the VC management policy as two separated aspects of the packet routing mechanism. In this way, the impact on performance of each of these two features can be separately studied in detail. The Valiant routing scheme has been selected to illustrate this analysis. As said before, Valiant has been used in different parallel machines regardless their topology and constitute the basis of other popular source-based non-minimal adaptive algorithms such as UGAL~\cite{Singh}. Moreover, other in-transit adaptive routing mechanisms, which employ non-minimal paths as in Valiant, are also strongly affected by the VC management policy, as the next example illustrates.

    Leveraging on DOR, the in-transit non-minimal adaptive DAL routing was introduced for HX in~\cite{HyperX}. In DAL, a deroute is allowed per dimension and packets can take up to $2n=2D$ hops. Buffers are divided into two sets. The adaptive subset is freely used, whereas the other subset constitutes an escape subnetwork, managed under DOR, that is used only when adaptive routes are blocked. A decade later, another in-transit non-minimal adaptive routing mechanism, denoted as OmiWAR, was proposed for HX in~\cite{Kim_omni}. It employs exactly the same paths as DAL, but instead of using an escape DOR virtual subnetwork for breaking deadlocks, OmniWAR employs the Ladder mechanism, which provides clearly superior performance.

    Scenarios as the previous one motivated the analysis of the two VC management previously introduced, \textit{2Phases} and \textit{Ladder}, including some variants.

    In the 2Phases mechanism, as it will be shown, the VC management at the intermediate switch, \IS{}, which signals the change of phase in the path, is critical in determining the performance of Valiant. A supposedly minor decision, as what to do in the cases in which either the source or destination switch is selected as intermediate, has a significant impact. When this occurs, a sensible solution that adequately balances traffic is to send these packets trough minimal routes, just needing one phase of the Valiant path. In these cases, it is needed to decide whether to inject them through the first or the second VC phase. We denote such policies as \textit{MinFirst} when minimally routed traffic uses the first VC phase, and \textit{MinLast} in the opposite case of injecting them trough the second phase.  A natural intermediate alternative could be \textit{MinBoth}, where packets would be injected through the first phase if the source switch is selected as the intermediate, and through the second if the intermediate is the destination switch. However, we will not consider this strategy in detail because it does not provide benefits, as it will be shown later on the paper.

    It should be noted that another policy for the cases in which source or destination switches are selected as intermediates, would be to obtain a new random \IS{} until neither the source nor the destination switch are obtained as intermediate. The performance of MinLast, when discarding source and destination from the pool of intermediate switches, would be similar to the performance of MinFirst when including them as intermediates, and the corresponding packets are sent minimally. Although not used in this paper, packets whose intermediates are switches belonging either to the source or destination groups, but not the source or destination switches themselves, could be also minimally routed. The rational behind this would be to avoid unnecessary local hops.

\begin{figure}
    \centering
    \includegraphics[scale=0.297]{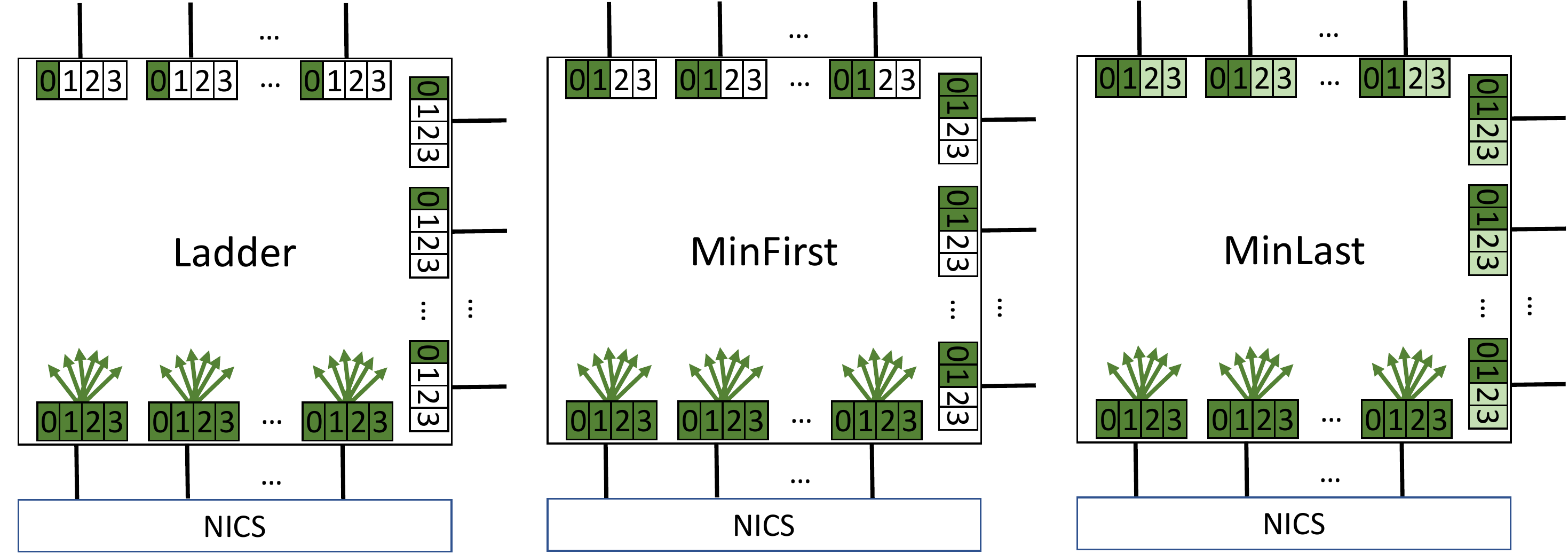}
	\caption{VCs allowed for injection depending on the VC management policy. Ladder injects at VC 0 (dark green); MinFirst inject at first phase (0--1, dark green); MinLast inject most commonly at first phase (0--1, dark green) and sporadically at the second phase (2--3, light green).}
	\label{fig:injection}
\end{figure}

    In~\cite{Kim_dgfly_ISCA}, when using UGAL for DFs, it is suggested to inject the minimal traffic through the second set of buffers, as with MinLast. But, as we will see, this is not a good choice, at least, with Valiant. Since only packets traversing the larger routes use the last VCs, intuitively, one could think that the buffers of the second phase are less demanded than those of the first one. Thus, injecting minimal traffic through the second phase would seem an adequate decision. The MinFirst mechanism, instead, injects all packets, including those using minimal paths, through the first VC phase. These differences at injection are illustrated in Figure~\ref{fig:injection}.


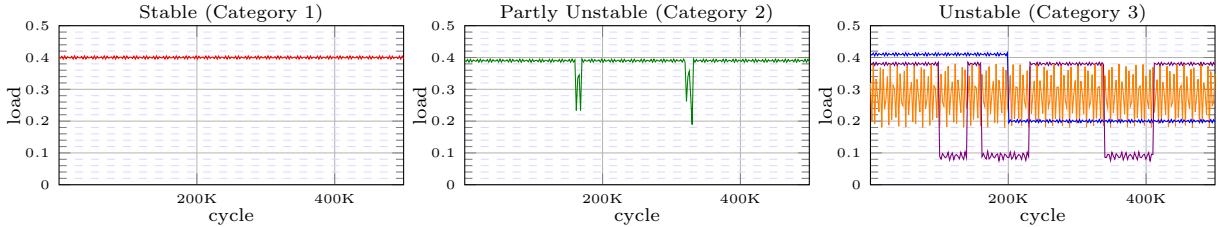
\begin{figure*}
	\centering%
	\tikzset{
		declare function={
			good(\x) = 0.4+0.005*sin (100*x);
			bad(\x)=(\x<=200)*(.41+0.005*sin(3+100*\x)) + (\x>200)*(0.2+0.005*sin(3+100*\x));
			carved(\x) = (.39+0.005*sin(5+100*\x))
				- and(\x>160, \x<=170) * 0.05*abs(\x-165)
				- and(\x>320, \x<=330) * 0.04*abs(\x-325);
			onoff(\x)= (.38+0.005*sin(7+100*\x))
				- or(or(and(\x>100,\x<140),and(\x>160,\x<230)),and(\x>340,\x<410)) * (.29+0.01*sin(20+120*\x));
			oscillating(\x)= .28+0.10*sin(24+103*\x);
		}
	}
    \tikzsetnextfilename{externalized-figure-categories-1}%
	\begin{tikzpicture}
	\begin{axis}[
        automatically generated axis,
		title={Stable (Category 1)},
        ymajorgrids=true,
        yminorgrids=true,
        xmajorgrids=true,
        minor y tick num=4,
		ymin=0,
		ymax=0.5,
		xmin=0,
		xmax=500,
		domain=0:500,
		samples=245,
		xtick={200,400},
		xticklabels={200K,400K},
		xlabel={cycle},
		ylabel={load},
		legend style={anchor=south,at={(0.5,1.05)},font=\scriptsize,fill=none},
	]
	\addplot[red,mark=none] {good(x)};
	\end{axis}
	\end{tikzpicture}%
    \tikzsetnextfilename{externalized-figure-categories-2}%
	\begin{tikzpicture}
	\begin{axis}[
        automatically generated axis,
		title={Partly Unstable (Category 2)},
        ymajorgrids=true,
        yminorgrids=true,
        xmajorgrids=true,
        minor y tick num=4,
		ymin=0,
		ymax=0.5,
		xmin=0,
		xmax=500,
		domain=0:500,
		samples=245,
		xtick={200,400},
		xticklabels={200K,400K},
		xlabel={cycle},
		ylabel={load},
		legend style={anchor=south,at={(0.5,1.05)},font=\scriptsize,fill=none},
	]
	\addplot[green!50!black] {carved(x)};
	\end{axis}
	\end{tikzpicture}%
    \tikzsetnextfilename{externalized-figure-categories-3}%
	\begin{tikzpicture}
	\begin{axis}[
        automatically generated axis,
		title={Unstable (Category 3)},
        ymajorgrids=true,
        yminorgrids=true,
        xmajorgrids=true,
        minor y tick num=4,
		ymin=0,
		ymax=0.5,
		xmin=0,
		xmax=500,
		domain=0:500,
		samples=245,
		xtick={200,400},
		xticklabels={200K,400K},
		xlabel={cycle},
		ylabel={load},
		legend style={anchor=south,at={(0.5,1.15)},font=\scriptsize,fill=none},
	]
	\addplot[orange] {oscillating(x)};
	\addplot[blue] {bad(x)};
	\addplot[violet] {onoff(x)};
	\end{axis}
	\end{tikzpicture}%
	\caption{Crafted examples of accepted load for each of the three Categories of stability.}
	\label{fig:categories}
\end{figure*}

    When combining Valiant routing with the previous VC management policies, the performance exhibited by the networks tested in this paper can be roughly classified in three categories. Figure~\ref{fig:categories} accompanies the following description, with a illustration of load lines resembling the results shown later.

     \textit{Category 1} would correspond to the desirable situation of a high accepted load, or throughput, which remains constant over time. Schemes in this category work well by themselves, without the need to add congestion control mechanisms. Solutions in this category can be directly compared by just looking the values of accepted load and latency achieved.

	\textit{Category 3} is characterized by instabilities that cause a notable drop in performance. In Figure~\ref{fig:categories} this category is illustrated with three different lines, each with a different pathological behaviour. Some may behave well until some point in time, when an abrupt drop occurs. Different simulations may have this drop in different points, and their average could suggest the decay is progressive instead of abrupt. Other possible behaviours are to irregularly fall and recover or to quickly oscillate between high and low values. Any of these behaviours is very undesirable, and the miscreant resulting routing could only be used along some additional congestion control mechanism.

    \textit{Category 2} has performance oscillations similar to those in \textit{Category 3}, but the drops
    in accepted load are followed by a recovery of the highest value.
	Different simulations will experience these drops at different points in time, and averaging a few simulations could look like a \textit{Category 1} scheme. However, these mechanisms should be treated with the same precautions as those in \textit{Category 3}, as it could be eventually a big enough drop from which the network cannot be recovered.

    It is important to note that some design decisions on the VC management, usually considered minor, can compromise the network performance. It means that is not difficult for a routing scheme to fall in \textit{Categories 2 and 3}. In these categories, the scenario is not as catastrophic as when having a deadlock, but it can seriously degrade performance and, what can be worse, to make the network response unpredictable.

\section{Experimental Setup}\label{sec:setup}



\begin{table*}[tb]
	\centering%
	\caption{The topologies employed in the evaluation.}
	\begin{tabular}{|l|l|l|l|l|l|l|}
	\hline
	\bf Topology & \bf Servers per switch & \bf Total servers & \bf Radix & \bf Switches & \bf Switch--switch links & \bf Minimal route used\\
	\hline
    1D HyperX $K_{64}$ & 64 & 4\,096 & 127 & 64 &4\,032 & single link \\
	\hline
	2D HyperX side 16 &16 & 4\,096 & 46 & 256 & 3\,840 & \texttt{XY}\\
	\hline
	3D HyperX side 8 & 8 & 4\,096 & 29 & 512 & 5\,376 & \texttt{XYZ}\\
	\hline
	Dragonfly $h=6$ & 6& 5\,256 & 23 & 876 & 7\,446 & \texttt{lgl}\\
	\hline
    Dragonfly+ $h=8$ & 8 (per leaf switch)& 4\,160 & 16 & 1\,040 & 6\,240 & \texttt{ugd}\\
	\hline
	\end{tabular}
	\label{tbl:topologies}
\end{table*}

Experiments carried out in this work have been performed on five different topologies: three HX with different number of dimensions, a DF, and a DF+. Each topology has been designed to accommodate around 5\,000 servers, and a summary of their parameters can be found at Table~\ref{tbl:topologies}.

%

As said before, a $n$D HX of side $s$ is the Cartesian power $K_{s}^{n}$. All the tested HX are able to equip 4\,096 servers. For the 1D case ($64^{2}$ servers), we employ 64 switches of radix 127, with 64 ports devoted to servers and the other 63, to connect to the other switches. The 2D HX uses 256 ($16^{2}$) switches of radix 46, with 16 ports for servers and 15 ports per dimension to connect to other switches. Finally, the 3D HX uses 512 ($8^{3}$) switches of radix 29, with 8 ports for servers and 7 ports per dimension for connecting to other switches. In all the cases, each of the two Valiant phases is performed under the deadlock-free DOR mechanism. Even when employing the Ladder policies, which do not require further deadlock preventions, DOR has been chosen to ensure a fair comparison.

The DF under test is a two-level hierarchical network composed of 73 groups connected as complete graphs of 12 switches. The global network is a complete graph of 73 groups, the maximum possible for the employed radix, so there is a single global link between any pair of groups. The switch radix is 23, with 6 ports devoted to servers, 11 ports to connect the other switches in the same group, and 6 ports to switches in other groups. The global links are arranged in the so called \textit{Palmtree} configuration~\cite{TACO}. The total number of switches is 876, which accommodate 5\,256 servers.
Its minimal paths, as detailed before, have at most 3 hops, are always unique, and have the form \texttt{lgl}. Nevertheless, the longest Valiant path will have the form \texttt{lgl-lgl}, with the intermediate switch (at the middle) marking the phase change. As said before, 2 local VCs and 1 global VC are required per phase to avoid deadlocks, giving a total of 4 VCs per local port and 2 VCs per global port. Another way to see this scheme is as two separated ladders, one with 4 steps for local ports and the other with 2, for global ones.

The DF+ is similar to the DF, but its groups are wired as complete bipartite graphs (leaf/spine Fat-trees) instead of complete graphs. In each group there are 8 leaf and 8 spine switches, each leaf connecting to each spine. In the first level, each leaf switch uses 8 ports to servers and 8 ports up to spine switches. In the second level, each spine switch uses 8 ports down to the leaf switches and 8 global ports towards other spine switches in remote groups. All the links between spines are between different groups and the resulting global network is the complete graph $K_{{65}}$. Thus, the network is composed of 65 groups, which accommodate 4\,160 servers in total, employing 1\,040 switches of radix 16. Alike in the DF, there are several possibilities for connecting the global links to make a complete graph, and we again employ the Palmtree configuration. In the DF+, minimal routes are up--down when intra-group and up--global--down when inter group, respectively written \texttt{ud} and \texttt{ugd}. Hence, they are deadlock-free when using minimal paths without the need of VCs.
The Valiant paths take the form \texttt{ugd-ugd}, which makes clear that 2 VCs per port are enough. This should be compared with the \texttt{lgl-lgl} paths in the DF, where the lack of subclasses of local links make necessary the use of 4 VCs for local ports.
Another way to see this scheme is as three separated ladders, one for local up ports, other for local down ports and the third for global ports, all of them with 2 steps.

Different policies for the selection of the intermediate switch have been used in Valiant routing, which could affect performance. However, in our simulations, any switch of the network is randomly selected as intermediate, in the cases of HX and DF. In DF+, any random leaf switch can be selected. This leads to routes whose length is limited to a maximum of 2D hops in all the tested topologies.

With respect to the load to which the different networks are exposed, we use both benign and adversarial traffic patterns. All the topologies have been dimensioned to accept load around the theoretical rate of 1.0 (one phit of a packet per server per cycle) when dealing with uniform traffic and minimal routing is used. However, performance can go up to a rate of 0.5 phits per server per cycle when non-minimal Valiant routing is used. Since in this work only Valiant routing is evaluated, every traffic pattern is limited to this 0.5 rate bound.

For benign traffic patterns, such as the uniform, that evenly distributes the load across all network resources, the different VC management mechanisms under study exhibit Category~1 performance under Valiant routing. Nevertheless,  Valiant should not be used for benign traffic, as minimal routing would give a better performance. Although out of the scope of this work, it should be noted that the different VC managers can exhibit different throughput values when managing uniform load under minimal routing.
For example, with Valiant in the HX, the Ladder shows lower performance than 2Phases, but this could be improved by removing the DOR restriction.
And the same occurs with minimal Ladder compared with DOR JSQ.
These details should be taken into account when designing source-based adaptive routings such as UGAL.
But, as stated before, our study on the potential unstable behaviour of the different VC managers focuses just on adversarial traffic patterns that really need Valiant load balancing.

A set of typical adverse traffic patterns has been chosen to evaluate the different VC management mechanisms in each topology. For the HX, we chose to show results of a ${\mathcal D}$\nobreakdash-$i$\nobreakdash-shift permutation traffic pattern in every dimension. The $\mathcal D$ being the set of dimensions in which the constant $i$-shift of switches is applied. For example, in Figure~\ref{fig:all_dimensions_hyperX}, the XY-7-shift means a shift of 7 switches in the X dimension, and 7 switches in the Y dimension.
That is, the servers connected to the switch labeled in the network as (0,0) send traffic to the servers connected to the switch labeled as (7,7).
In general, the servers of the switch $(x, y)$ send traffic to the servers of the switch $(x+ i, y+i)\mod s$, where $s=16$ is the side of the 2D HX taken as example. For both Dragonflies, a commonly employed adverse traffic pattern is ADV+i, where each server at a group $G$ sends traffic to the server located at the same relative position at the $G+i$ group.
Taking $i$ as the number of global links used per switch, which is denoted as $h$, is most adverse for certain DF configurations~\cite{Prisacari}, and we have selected it.
We also showcase a variant of ADV+h denoted as ADVr+h, where traffic is directed to a random server in the target group. The ADVr+i pattern has been extensively used in the literature, but it appears that is less adverse than the ADV+i pattern.

The number of VCs employed for each mechanism and topology is specified in Table~\ref{tab:vc}. To ensure a fair comparison, all the VC management mechanisms were allocated the same resources. However, in the case of the DF, the 2Phases managers were only allowed to have either 2/4 or 4/8 (global/local) VCs, while the Ladder mechanism needs just 6 VCs. Nonetheless, it was observed that this difference did not affect performance significantly, so the presented results are based on the 4/8 configuration.

\ifarxiv
To conduct the experiments, we employed a public in-house developed network simulator CAMINOS~\cite{CAMINOS}.
\fi
When traffic is adversarial and offered load is above saturation, we observe in our tests diverging performance results for 60-80\% between the different VC mechanisms. These divergences indicate that the observed effects are not limited to few corner cases. Furthermore, previous studies have acknowledged the existence of similar effects in Valiant and other non-minimal routing schemes ~\cite{benito2019acor, Kim_omni}, using other comparable network simulators, such as Booksim2.0~\cite{booksim} and SuperSim~\cite{mcdonald2018supersim}, although they did not specifically focus on studying them as extensively as we do in this work.

We simulate a typical router model, with FIFO buffers at both input and output ports and a basic allocator. Other parameters used in the simulation are gathered in Table~\ref{tbl:extra_parameters}. A duration of 510k cycles has been empirically chosen as a value large enough to encompass most of the observed instabilities.
It is worth noting that these simulation runs are larger than the typical convergence time, casting some doubt on whether there are potential instabilities on previous studies with shorter times.

\begin{table}
    \centering%
    \caption{Parameters in the simulation.}
    \begin{tabular}{|l|l|}
        \hline
        \bf Parameters & \bf Value \\
		\hline
        Simulated cycles & 510k \\
		\hline
        Input buffer size & 4 packets \\
		\hline
        Output buffer size & 2 packets \\
		\hline
        Flow control & Virtual cut-through \\
		\hline
        Packet length & 16 phits \\
        \hline
    \end{tabular}
    \label{tbl:extra_parameters}
\end{table}

Two types of charts have been generated to visualize the results. The first represents a scale-up scenario, where the offered load increases, and the average accepted load, or throughput, is plotted for each offered load level. The second type of chart displays dynamic temporal results, showing the accepted network load averaged in bins of 1K cycles. In these temporal simulations, the offered load parameter is always set to its maximum value of 1.0. The plotted results are averaged over ten runs, each run with a different seed of the random number generator. In some cases, we observed important differences between runs, and show figures with a line for each one of the ten simulations, instead of averaging them. Indeed, in these highly unstable scenarios, the average can result very misleading for a reader unaware of the phenomenon.




\begin{table*}[htbp]
    \centering
    \caption{Number of VCs used per strategy in each topology.}
    \begin{tabular}{|l|c|c|c|c|}
        \hline
        \textbf{VC Managment} & \textbf{2D HyperX} & \textbf{Dragonfly} & \textbf{DragonflyF+} & \textbf{3D HyperX}  \\
        \hline
        2phases MinFirst & 2 vcs + 2vcs & 2/4 vcs + 2/4 vcs & 3vcs + 3vcs & 3vcs + 3vcs \\
        \hline
        2phases MinLast & 2 vcs + 2vcs & 2/4 vcs + 2/4 vcs & 3vcs + 3vcs & 3vcs + 3vcs \\
        \hline
        Ladder & 4vcs & 6vcs & 6vcs & 6vcs \\
        \hline
        Ladder with reuse & 4vcs & 6vcs & 6vcs & 6vcs  \\
        \hline
    \end{tabular}%
    \label{tab:vc}%
\end{table*}%

\section{Results}\label{sec:results}

Many simulations have been conducted to evaluate the performance of the different VC management policies, varying topologies, traffic patterns, and switch configurations. A selected subset of results, considered representative enough to draw conclusions, is presented next.  The VC management strategies that will be assessed are listed in Table~\ref{tab:vc}.



Figure~\ref{fig:all_minb_temporal} depicts the core problem addressed in this paper by showing the accepted load of the 2Phases MinLast mechanism across the three topologies under adversarial traffic patterns, for a 1.0 offered load parameter (each server generating at full rate, 1 phit per cycle). Each chart displays the accepted load, or throughput, at each cycle, averaging 10 runs. Remember that the maximum achievable throughput for Valiant is 0.5. The figure clearly illustrates the performance instabilities caused by MinLast. All three charts fall under Category~3. In the case of the 2D HX, 6 out of the 10 simulations experienced a decline from approximately 0.4 to 0.04, without recovery. For the DF, the throughput oscillates between the 0.4 and 0.2 range, while in the case of the DF+, the throughput is distributed between 0.4 and 0.04.

The MinBoth mechanism, whose behavior is depicted in Figure~\ref{fig:minBoth}, shows similar but more alleviated results than MinLast. However, as we said, it is not going to be further studied because it does not add significant data and its performance is dependent on the MinLast mechanism.

\input{figures/Seeds/all_minb}
\input{figures/minBoth}

Next, performance results are presented and analyzed per topology, starting with HX networks, from one to three dimensions, followed by DF networks, and finally, DF+ networks.


\subsection{HyperX (HX)}


The performance of the VC managers was evaluated across one to three dimensional networks, using a shift switch permutation traffic pattern. The results can be seen in Figure \ref{fig:all_dimensions_hyperX}.

In a 1D HX there were no significant differences observed between the VC policies in both the scaling-up and temporal simulation charts. All mechanisms exhibited Category 1 performance, indicating high throughput and stability. In this case, the Ladder mechanism is equivalent to MinFirst, so it is not shown separately.

In a 2D HX, congestion issues arose when applying an XY-7-shift traffic pattern. The MinLast mechanism showed Category 3 performance with significant instability after saturation, as evident in the upper and lower charts. Specifically, the temporal chart with the MinLast mechanism was previously shown without averaging the 10 runs in Figure~\ref{fig:all_minb_temporal}, which further demonstrates the instability and variance among simulations. It appears that the performance will not recover, and all simulations would eventually decline to a performance level around 0.04 after enough cycles. Notably, the temporal simulations showed a significant performance drop after 200K cycles, which would have gone unnoticed in shorter runs. On the other hand, MinFirst maintained the performance above 0.4 for most of the simulation, with a slight decline that later recovered and remained stable. However, since this sporadic drop could occur at any time in a simulation with more cycles, the overall stability of MinFirst is not clear. Therefore, in a pessimistic classification, we could consider the performance of MinFirst as Category~2 in this case. The Ladder mechanism, as it is represented in the charts, provides clear Category~1 performance. Nevertheless, in average, the throughput is lower than MinFirst. The Ladder with reuse would obtain the best performance overall, being equal in average than MinFirst and providing stable results as the basic Ladder. Thus, among the Category~1 mechanisms for this network, the Ladder with reuse is a very reliable policy and a clear winner.

Finally, in a 3D HX applying a XYZ-4-shift traffic pattern, both 2Phases mechanisms fall early under Category 3 performance, with an average accepted load around 0.2. On the contrary, the Ladder mechanism maintains stability and shows throughput values slightly below 0.4 phits per server per cycle, so can be classified as Category 1 performance. Therefore, to achieve a stable performance in 3D HX networks, the use of VCs should be limited to the Ladder mechanism. Furthermore, the Ladder can be enhanced by utilizing VC reuse as in the 2D HX, achieving an average accepted load above 0.4, which could be the desired level of performance. 

\input{figures/HyperX_general/hyperx_dimensions}

These findings suggest that the Ladder is a robust solution that can provide stable performance across HX networks of different dimensions and traffic conditions. Furthermore, the Ladder mechanism with VC reuse could be the best solution for these networks. On the other hand, the performance of Valiant 2Phases variants appears to be more sensitive to the network topology and traffic patterns. Apparently, MinLast only achieves Category 1 performance in a 1D HX, while MinFirst is able to provide an outstanding Category~2 performance for a 2D HX. However, both obtain a Category~3 performance in a 3D HX. This highlights the importance of carefully considering these insights when selecting a VC management policy for high-dimensional HX networks.

\subsection{Dragonfly (DF)}

\input{figures/Dragonfly_general/dragonfly_general}
The evaluations in the DF network includes the use of two traffic patterns: ADVr+h and ADV+h. The performance results for this network are depicted in Figure~\ref{fig:dragonfly_patterns_performance}.

In the case of the ADVr+h traffic pattern, it can be observed that MinLast behaves badly. Both scale-up and temporal charts prove the degradation of performance. In the scale-up chart, throughput starts to decline after saturation, resulting in a loss of the 0.4 level. The temporal chart shows that throughput eventually decreases to a value of 0.24, and it is doubtful whether it could decline further. Each run in the temporal simulation of this experiment with MinLast is plotted independently in Figure~\ref{fig:all_minb_temporal}, further emphasizing its instability and poor performance. Based on these results, MinLast is classified as Category 3 performance. On the other hand, MinFirst maintains a stable performance above 0.4 throughout the entire simulation, exhibiting Category 1 performance. The Ladder shows a stable performance throughout below 0.4, which is lower than the MinFirst performance. However, the Ladder with reused VCs matches the performance of MinFirst, achieving a level of accepeted load around 0.44, as shown in both the temporal and scale-up charts.

For the ADV+h traffic pattern, it can be observed that both 2Phases mechanisms experience congestion problems when injecting traffic beyond saturation. It is worth noting that MinFirst is clearly degraded after 200K cycles of simulation, which could not be appreciated if it were shorter. In contrast, the Ladder exhibits stable performance slightly below 0.4 the entire simulation. Hence, it is classified as Category~1. Similarly, the Ladder with reused VCs achieves a constant performance level of around 0.44, as can be seen in both the temporal and scale-up charts. Therefore, the Ladder with reuse is clearly Category~1. These results indicate that the Ladder, especially when enhanced with VC reuse, provides stable and high performance in the DF network under adverse traffic patterns.

In conclusion, the evaluation of the VC managers in the DF network, when dealing with the ADVr+h and ADV+h traffic patterns, revealed distinct performance characteristics. However, it is not usually to show the difference between these two patterns. The MinLast mechanism always exhibited lower performance and notable instability, ranking it as Category 3. In contrast,  MinFirst proved stable Category 1 performance in ADVr+h, but Category~3 in ADV+h. Both Ladders are always Category~1 performance, being the Ladder with reused VCs the best option. These findings emphasize the effectiveness of the Ladder mechanism, particularly when combined with VC reuse, in providing stable and high-performance solutions in the DF.

\subsection{Dragonfly+ (DF+)}

\input{figures/Dragonfly+_general/Dragonfly+_general}

The results of the evaluation of the DF+ network, again using the two traffic patterns ADVr+h and ADV+, are depicted in Figure~\ref{fig:DF+_comparison_general}.

Under the ADVr+h traffic pattern, MinLast exhibited Category 3 performance, experiencing congestion problems after saturation. Each independent run in the temporal simulation of the MinLast mechanism is presented separately in Figure~\ref{fig:all_minb_temporal}, where the dependency of the instability on the specific run is evident. In contrast, MinFirst maintained a stable performance above 0.4, indicating Category 1 performance. The Ladder also demonstrated Category 1 performance but with a lower throughput compared to MinFirst. Interestingly, the Ladder with reused VCs provided also lower performance than MinFirst.

The ADV+h pattern also affects MinLast in the same way than ADVr+h, resulting in Category~3 performance. Notwithstanding, MinFirst exhibited Category 1. This difference is notable when compared to the results observed in the DF network, where MinFirst did not sustain the desired performance level. Even in 2D and 3D HX networks, MinFirst could not sustain the performance registered for the DF+. The Ladder and Ladder with reused VCs also are Category~1, but the accepted load is lower than in the MinFirst mechanism. The Ladder is far from the performance of MinFirst but the Ladder with reused VCs has a slightly lower performance.

Overall, the MinFirst mechanism in the DF+ consistently demonstrated Category 1 performance, while the Ladder mechanism provided a stable solution but with a lower throughput. The Ladder with reused VCs had a similar performance level to MinFirst, but slightly lower.

%

\subsection{Discussion}

 As it has been seen, different networks with different VC managers suffer from congestion. Such performance degradation after saturation is always due to HoLB issues caused by congestion trees. It is known that using one VC per end-point destination, as in the Avici router, eliminates any HoLB at network level~\cite{Dally}. This solution constitutes an extreme point in cost. Alternatives, such as virtual output queuing (VOQ), uses as many VCs as the switch radix and fully eliminates HoLB at switch level. With high-radix switches, as the ones used in current networks, VOQ is still prohibitively expensive.

 Both 2Phases and Ladder mechanisms are very low-cost in terms of VCs. Hence, both can be potentially considered congestion-prone, but as they use buffers in a different manner, their behavior is quite different across networks.
As a general conclusion, injecting minimal traffic trough the second VC phase, as done by MinLast in the 2Phases mechanism, is a bad design decision.
Most of the traffic in MinLast is injected through the first VC phase, but the few packets injected through the second phase cause interferences to the in-transit packets, which exacerbates HoLB issues that impede the delivering of packets to destinations. Thus, MinLast should be totaly avoided in almost any network.

 The 2Phases policy uses 2 or 3 VCs per phase so it can segregate traffic into 2 or 3 flows and, in this way, to reduce some HoLB in proportion to the number of the VCs used. Nevertheless, the Ladder reduces HoLB by segregating flows according to the distance traversed by packets. Remember that a Ladder uses $2D$ steps, with $D$ the network diameter. Hence, the Ladder segregates traffic into $2D$ classes, reducing also HoLB in proportion to the segregated classes. Nevertheless, the Ladder offers less buffer room per hop, as it uses just one VC per step, while 2Phases uses 2 or 3 per phase. Both VC managers guaranty deadlock freedom but exhibit different levels of HoLB depending on the network. HX and DF behaves more or less the same, but DF+ is different.

 To simplify, focussing on DF and HX networks, MinFirst is unable to avoid congestion and instability except for the 1D HX. The case of this complete graph is special, as distances are too short and minimal traffic takes just one hop, making the VC competition almost non-existent. Nevertheless, for the 2D and 3D cases, the segregation of flows per traversed distance of the Ladder appears much more powerful in alleviating HoLB than 2Phases, that uses several VCs per phase under a JSQ discipline.

 The situation is different with DF+. Due to the acyclic nature of its groups, the DF+ is simpler than the other two networks. The are more classes of packets, with their corresponding lengths, in the DF and HX networks than in the DF+. In our simulations we have just four classes of paths in a DF+: \texttt{ugd}, \texttt{ud-ugd}, \texttt{ugd-ud} and \texttt{ugd-ugd}. In the other networks, the number of path classes and their variability on length is quite superior, which leads to higher levels of packet contention and HoLB that only a Ladder is able to overcome. As it has been seen, the most adequate VC management mechanism for DF+ networks is 2Phases with MinFirst. In this scenario, is more efficient for mitigating HoLB the use several VCs per phase in 2Phases than a Ladder with just one VC per step. Adding VCs to the Ladder steps will increase its performance but also its cost.

\section{Conclusions}\label{sec:conclusions}

In this study, we conducted a comprehensive evaluation of the VC management policies that can be applied to modern network topologies, including HyperX, Dragonfly, and Dragonfly+. The performance of these mechanisms was analyzed under various traffic patterns, aiming to identify the most effective solutions.

Our findings highlight the importance of selecting the appropriate VC mechanism for a given network architecture and traffic scenario. On the one hand, it has been shown that 2Phases should not be use neither in HyperX nor in Dragonfly networks. MinLast exhibited very poor performance, making it an unreliable choice for any scenario. At the same time, the MinFirst mechanism showed a more consistent performance in some cases, but not totally stable in general. For such networks a Ladder with reused VCs is the clear winner mechanism.

On the other hand, in the Dragonfly+, among all the mechanisms tested, 2Phases MinFirst has consistently shown the best performance. This finding is noteworthy as MinFirst exhibited lower performance than a Ladder in both Dragonfly and HyperX networks. The results suggest that MinFirst is a reliable and effective solution for high-performance communications in the Dragonfly+ network.

Interestingly, the Ladder mechanism, although providing stable performance under any circumstance, demonstrated lower throughput compared to MinFirst in some cases. However, the Ladder with reused VCs achieved almost the highest stable performance overall, making it an appealing option when stability and high-performance are paramount.


Future research will focus on refining the mechanisms presented in this paper, and test them when integrated with either source-based or in-transit adaptive non-minimal routing schemes.



\section*{Acknowledgment}

\ifarxiv
The second author is under Ramón y Cajal contract RYC2021-033959-I from Spain's Ministerio de Ciencia e Innovación with funding from the Mecanismo de Recuperación y Resiliencia de la Unión Europea.
The four authors participate in these projects: PLANIFICADORES Y REDES PARA DATA CENTERS SOSTENIBLES project TED2021-131176B-I00 with funding from MCIN/ AEI /10.13039/501100011033 and Unión Europea/NextGenerationEU/PRTR; and REDES DE INTERCONEXIÓN, ACELERADORES HARDWARE Y OPTIMIZACIÓN DE APLICACIONES, project PID2019-105660RB-C22 with funding from MCIN/ AEI /10.13039/501100011033.

In addition, the first author thanks to the Quantum Spain consortium and the fourth, to the BSC and Nvidia for supporting their research.

Some of the simulations have been performed in the Altamira supercomputer, of the Spanish Supercomputing Network (RES), located at the Institute
of Physics of Cantabria (IFCA).


\fi

\ifarxiv
\bibliographystyle{plain}
\fi
\bibliography{main}

\end{document}